# Liquid crystalline structures formed by sphere-rod amphiphilic molecules in solvents.

Nilanthi P. Haputhanthrige[1,2], Yifan Zhou[3], Jingfan Wei[3], Min Gao[1], Tianbo Liu[3], and Oleg D. Lavrentovich[1,2*]

[1] *Advanced Materials and Liquid Crystal Institute, Kent State University, Kent, OH 44242, USA*
[2] *Department of Physics, Kent State University, Kent, OH 44242, USA*
[3] *School of Polymer Science and Polymer Engineering, The University of Akron, Akron, OH 44325, USA*

***Correspondence**: olavrent@kent.edu

**Abstract**

Self-assembly of amphiphilic molecules is an important phenomenon attracting a broad range of research. In this work, we study the self-assembly of $KTOF_4$ sphere-rod amphiphilic molecules in mixed water-dioxane solvents. The molecules are of a T-shaped geometry, comprised of a hydrophilic spherical Keggin-type cluster attached by a flexible bridge to the center of a hydrophobic rod-like oligodialkylfluorene (OF), which consists of four OF units. Transmission electron microscopy (TEM) uncovers self-assembled spherical structures of $KTOF_4$ in dilute solutions. These spheres are filled with smectic-like layers of $KTOF_4$ separated by layers of the solution. There are two types of layer packings: (i) concentric spheres and (ii) flat layers. The concentric spheres form when the dioxane volume fraction in the solution is 35-50 vol%. The flat layers are formed when the dioxane volume fraction is either below (20 and 30 vol%.) or above (55 and 60 vol%.) the indicated range. The layered structures show no in-plane orientational order and thus resemble thermotropic smectic A liquid crystals and their lyotropic analogs. The layered packings reveal edge and screw dislocations. Evaporation of the solvent produces a bulk birefringent liquid crystal phase with textures resembling the ones of uniaxial nematic liquid crystals. These findings demonstrate that sphere-rod molecules produce a variety of self-assembled structures that are controlled by the solvent properties.

# 1. Introduction

Self-assembly of amphiphilic molecules in aqueous solutions draws significant attention in many fields, such as molecular chemistry, biochemistry, and biomedicine [1-6]. The self-assembled structures are diverse, ranging from isolated spherical micelles and vesicles to helices, fibrils, and extended liquid crystals (LCs) such as lyotropic $L_\alpha$, $L_1$, and $L_3$ phases [7-22]. Lyotropic LCs, formed by the self-assembly of amphiphilic molecules, have received considerable attention due to diverse research areas such as drug delivery [6,23-27], and molecular engineering [28-30]. Self-assembly of amphiphilic surfactants and flexible block copolymers have been well studied with the rule of "packing parameters", which predicts the formation of micelles, wormlike micelles, vesicles and bilayers, etc. However, when some components, especially the solvophobic ones, are rigid and cannot comply with the conformational changes during the assembly, the common rules will not be applicable. An interesting example was recently introduced by Luo et al. [13], who studied amphiphilic $KTOF_4$ molecules in the shape of a letter T, in which a hydrophilic spherical Keggin-type cluster connects by a flexible link to a center of a hydrophobic rod-like oligodialkylfluorene (OF), Figure 1(a) [13]. In the formula of $KTOF_4$, "4" indicates the number of connected OFs, and "T" represents the shape of the sphere-rod molecule, Figure 1. The chemical formula of Keggin is $(XM_{12}O_{40})^{4-}$ where X is pentavalent phosphorus ($P^V$), M is tungsten (W) metal, and O is oxygen. The OF is a combination of carbons and hydrogens in the shape of a rod. The Keggin clusters and OFs are rigid, while the chain that connects them is flexible, thus allowing the Keggin and OF segments to move with respect to each other. The studies by Luo et al. [13] revealed that in water and acetonitrile (MeCN) mixed solutions, $KTOF_4$ molecules self-assemble into spherical inclusions with a system of concentric layers in their interior. The concentric spherical layers fill the entire volume of the inclusions with identical interlayer distance. The number of $KTOF_4$ bilayers in the concentric spherical inclusion, and thus the size of the inclusion, depends on the solvent polarity, temperature, and concentration [13]. In a solution with high polarity, 85 vol% water, the inclusion radius is about 19 nm including 3 concentric spherical bilayers. In a solution with lower polarity, 50 vol% water, the inclusions grow to a radius of 65 nm with 12 bilayers.

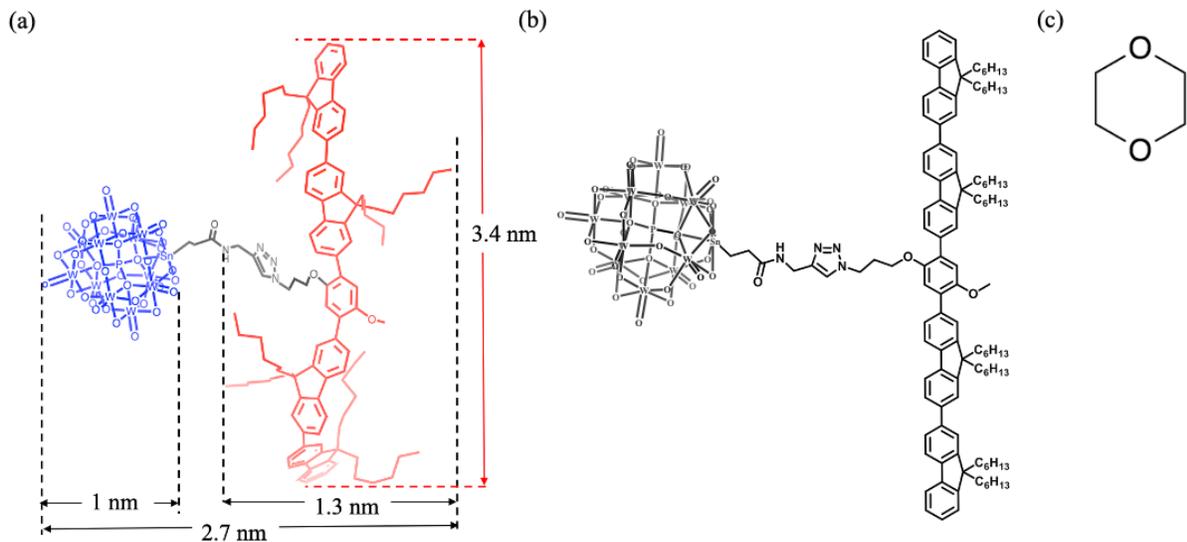

**Figure 1.** Chemical structures of materials. (a) 3D, (b) 2D representation of sphere−rod shaped KTOF$_4$ with the hydrophilic Keggin cluster attached to the center of OFs, and (c) dioxane. The width and the length of the rod-shaped OF are approximately 1.3 and 3.4 nm, respectively, and the diameter of the Keggin sphere is 1.0 nm. The Keggin cluster is indicated in blue, and the OF units are shown in red in Figure 1(a)

Spherical concentric packings of layers are frequently observed in droplets of thermotropic LCs such as smectics A (SmA) [31], smectics C (SmC) [32], and cholesterics [33-37]. Depending on the type of the order parameter, these spherical concentric packings of layers show two distinct geometries. In SmA, the orientational order is uniaxial, with the molecules oriented along the director $\hat{\mathbf{n}}$ which is along the normal $\hat{\mathbf{v}}$ to the layers. A spherical packing is then of a pure radial type, with $\hat{\mathbf{n}}$ along the radial directions, forming an isolated point defect-hedgehog in the center of the droplet. In SmC, the molecular tilt within the layers produces an additional degree of orientational order, described by a polar vector $\mathbf{c} = \hat{\mathbf{n}} - \hat{\mathbf{v}}(\hat{\mathbf{n}} \cdot \hat{\mathbf{v}})$ tangential to the layers; here $\hat{\mathbf{n}}$ is tilted with respect to $\hat{\mathbf{v}}$. As a result, the spherical packings still carry a radial point defect in the normal to the layers, but now this point defect is not isolated: it is connected to one or two disclination lines in the vector field $\mathbf{c}$ [32]. Similar "Dirac monopole" concentric packings with radial disclinations can be observed in cholesteric droplets [33-36,38-42].

Although the spheres of KTOF$_4$ inclusions show concentric packings, it is not clear whether the structures are of a SmA type, i.e., with no orientational order, or of a SmC/cholesteric type, i.e., with an orientational order within the layers. Such an orientational order might be expected thanks to the presence of the rod-like OFs, which might prefer to align parallel to each other and form a two-dimensional nematic ordering within the layers, described by a nonpolar version of the vector **c**. If this is the case, then the spherical KTOF$_4$ inclusions would be expected to show radial disclination lines in the **c** field.

The goal of this work is to explore whether the self-assembled layers of T-shaped KTOF$_4$ molecules have an isotropic or orientationally ordered in-plane structure, by analyzing the nanoscale-resolved transmission electron microscopy (TEM) images of their spherical inclusions in a mixture of water and 1,4-Dioxane (dioxane), Figure. 1(c). In solutions with 0.2 mg/ml KTOF$_4$, these spheres exhibit two main inner structures: (i) concentric spherical layers and (ii) flat parallel layers, Figure 2. In the latter case, we do not observe radial disclinations, which indicates that there is no orientational order within the layers, i.e., the KTOF$_4$ molecules show an isotropic in-plane ordering, similar to the structure of a SmA. The layered SmA structure shows edge and screw dislocations. The inner geometry of spherical inclusions is determined by the dioxane volume fraction in the solution. Namely, the solutions with a scarcity or excess of dioxane form flat layers, in which the concentration of dioxane in the range 35-50 vol% yields concentric spheres. The flat layers in polar solvents might be explained by their increased electrostatic repulsion between negatively charged Keggin clusters and by hydrophobic attraction of the rod-like OFs. Increase of the concentration of the KTOF$_4$ molecules through vaporization of the solvent can produce a homogeneous nematic-like bulk phase with high birefringence of 0.2.

## 2. Materials and Methods

### 2.1. The Material Synthesis

The alkyne-functionalized Keggin, $((PW_{11}O_{39})(SnCH_2CH_2CONHCH_2CCH))^{4-}$, with 4 tetrabutyl ammonium counterions (TBA$^+$) is synthesized based on a reported method in the literature [43], and the azide-containing OFs are synthesized through the step-by-step addition of fluorene repeating units [44,45]. Two components are coupled using azide−alkyne Huisgen cyclo-addition to form sphere−rod shaped hybrids KTOF$_4$, Figure 1(a), as explained in the ref [13]. Different 0.2 mg/ml KTOF$_4$ samples are prepared by changing the dioxane volume fraction from 20 to 60 vol% in the water and dioxane mixed solvents.

To estimate the dimensions of the KTOF$_4$ molecule, the molecular structure is drawn using ChemDraw software (version: 23.1.2). The 3D structure of an isolated molecule in a vacuum is obtained using MM2 Energy minimization engine of Chem3D software (Perkin Elmer, version: 20.0.0.41). Then, the dimensions of energy minimized molecule are measured in chem 3D and presented in Figure 1(a).

*2.2. Transmission Electron Microscopy.*

A small volume (~3 μl) of the sample solution is placed on a carbon coated copper grid and allowed to dry. The dried TEM grids are used for TEM analysis. Regular TEM images are captured using an FEI Tecnai F20 microscope (200 kV). The basic experimental setup and the procedure are detailed by Gao M. et al. [46]. The visibility of the layered structures in the TEM images is improved by enhancing the contrast and brightness using Fiji/ImageJ, version 2.140/1.54f.

*2.3. Polarized Optical Microscopy Study of High Concentrated KTOF$_4$.*

We separately study highly concentrated solutions of KTOF$_4$. A 2.0 mg/ml KTOF$_4$ in 50 vol% dioxane is placed in a capillary tube to allow the mixture to evaporate slowly and to increase the KTOF$_4$ concentration further. A new material with birefringent colors, abbreviated LC$_{KTOF4}$, forms at the capillary tube's meniscus. The textures of LC$_{KTOF4}$ sessile droplets at glass plates are explored under an Olympus BX51 polarized optical microscope (POM) (Olympus, Tokyo, Japan).

The birefringence of the LC$_{KTOF4}$ is measured in a wedge cell with two PI2555 polyimide-coated, rubbed glass plates as described in ref. [47,48]. The temperature of the cell is increased until LC$_{KTOF4}$ reaches the isotropic phase. The material is then cooled to the nematic phase (N) while observing under a POM with crossed polarizers. The temperature of the cell is controlled using a Linkam hot stage (Linkam Scientific, Redhill, UK).

## 3. Results

We prepare different samples of 0.2 mg/ml KTOF$_4$ by changing the dioxane to water volume fraction. Throughout the text, "X vol% dioxane" refers to a solution with a dioxane: water ratio of X:(100-X). Unless stated otherwise, a "KTOF$_4$ sample" refers to a sample containing 0.2 mg/ml KTOF$_4$. KTOF$_4$ solutions are studied using bright-field TEM textures. Self-assembled inclusions are observed across the dioxane content of 20-60 vol%. Precipitation occurs when dioxane content is below 20 vol%, and the self-assembly is not observed when dioxane exceeds 70 vol%.

KTOF$_4$ molecules aggregate into spherical inclusions with radii 25-90 nm, Figure 2(a-i). The inclusion size does not correlate with the dioxane volume fraction through the TEM results. As revealed in the TEM images, the assembled inclusions are filled with periodic structures exhibiting alternating dark and bright layers. Two distinct layered structures are observed in the TEM textures: one consists of concentric spherical layers, and the other consists of flat layers arranged inside a spherical shape. In the solutions with 20 and 30 vol% dioxane, the spherical structure with flat layers is dominant. As the dioxane volume fraction increases to 35-50 vol%, the concentric spherical layer structure becomes dominant. However, at higher dioxane volume fractions, 55 and 60 vol%, the flat layers are dominant again, Figure 2 & Table 1. The repeated interlayer distance remains constant at 4.8 nm and does not vary with the dioxane volume fraction. We are not aware of the reentrant packing scenario with a sequence flat-spherical-flat as a function of composition in any other lyotropic layered liquid crystal system.

The composition of the surface layer of the spherical inclusions varies with the dioxane volume fraction. In solutions with 20, 55, and 60 vol% dioxane, which form flat layers, some inclusions exhibit a dark interfacial layer with a thickness of 0.8-1.1 nm, which can be associated with the Keggin clusters, while others have a bright interfacial layer with a thickness of 1.4-1.9 nm, which might correspond to the combined extension of the flexible bridge and OF rods along the horizontal direction in Figure 1, which amounts to about 1.7 nm. A thick bright layer, with a thickness of 3.8-6.5 nm greater than that of a KTOF$_4$ molecule, is observed around some inclusions in compositions with 30, 35 & 55 vol% dioxane, Figure 3. These thick, amorphous layers might be a combination of KTOF$_4$ and TBA$^+$ ions that accumulate around the self-assembled spheres during drying. The layer beneath this thick bright periphery in 30, 35 & 55 vol% dioxane compositions is dark and thus corresponds to Keggin clusters, Figure 3. When the dioxane volume fractions are 40, 45, and 50 vol%, the surface layer is a bright thin layer made of OF rods. Note that the concentrations 35, 40, 45, and 50 vol% dioxane correspond to the spherical packing of layers. Since in these cases the interfacial layer is comprised fully of either OFs (for 40, 45, and 50 vol% dioxane) or Keggin clusters (35 vol% dioxane), we conclude that the spherical packing can be facilitated by the surface tension anisotropy, which prefers one component of the molecule to be at the interface. For flat layers, the preference is apparently weaker.

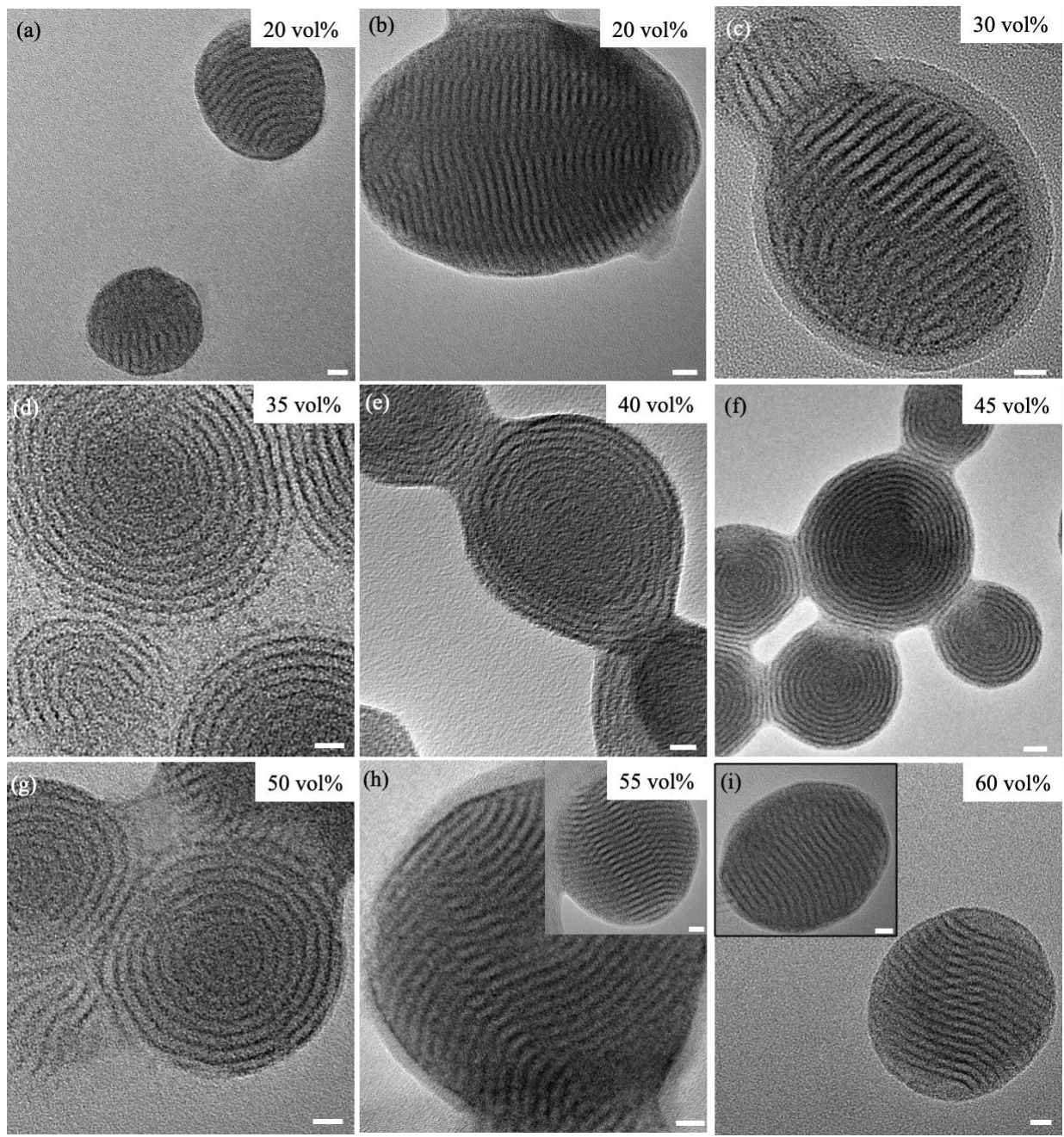

**Figure 2.** TEM textures of 0.2 mg/ml KTOF$_4$ in a dioxane and water mixture. The solution with (a & b) 20, (c) 30, (d) 35, (e) 40, (f) 45, (g) 50, (h) 55, and (i) 60 vol% dioxane. The scale bar is 10 nm. X vol% dioxane is a solution with a dioxane: water ratio of X:(100-X).

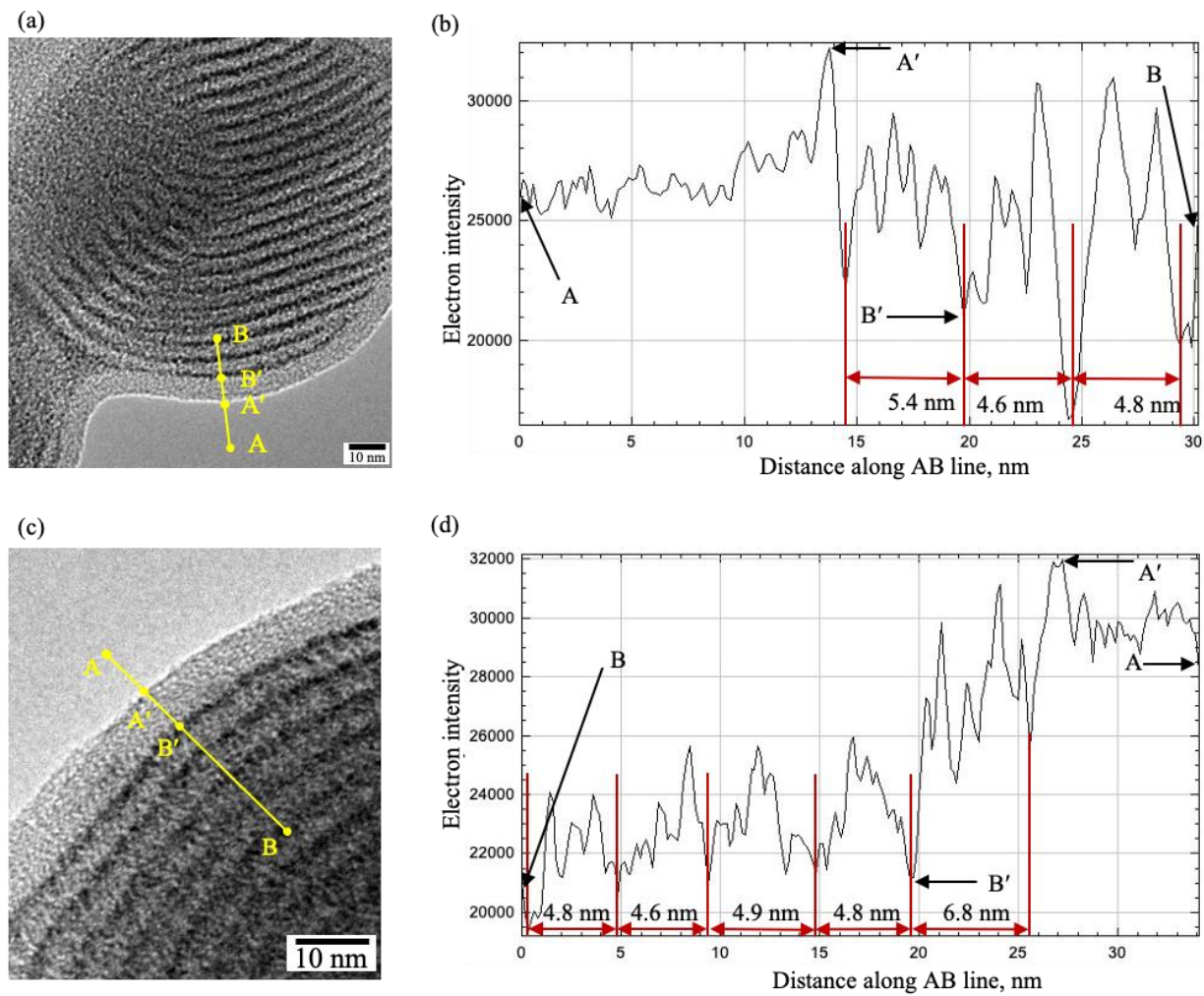

**Figure 3.** Self-Assembled inclusions with thick bright peripheral layer. TEM textures of 0.2 mg/ml KTOF$_4$ in a (a) 30 vol%, and (c) 35 vol% dioxane. (b & d) Transmitted electron intensity profile along the line drawn in (a) and (c) respectively. X vol% dioxane is a solution with a dioxane: water ratio of X:(100-X).

**Table 1**. Summary of layer type and surface layer composition of inclusions in solutions with different dioxane volume fractions. (X vol% dioxane; dioxane: water=X:(100-X))

| Dioxane vol% | Concentric layers | Flat layers | Surface composition |
|---|---|---|---|
| 20 | Not observed | Frequently observed | Keggin/OF |
| 30 | Not observed | Frequently observed | Keggin |

| 35 | Frequently observed | Not observed | Keggin |
| 40 | Frequently observed | Not observed | OF |
| 45 | Frequently observed | Not observed | OF |
| 50 | Frequently observed | Observed (but rare) | OF |
| 55 | Not observed | Frequently observed | Keggin/OF |
| 60 | Not observed | Frequently observed | Keggin/OF |

A zoomed-in TEM texture of the 30 vol% dioxane sample is used to study the assembled structure in detail. The bright-field TEM imaging condition used in this study produces textures mainly governed by mass-thickness contrast. Higher atomic number area generates more scattered electrons at higher angle, resulting in a reduced amount of transmitted electron density compared to molecules containing atoms with lower atomic numbers [49,50]. For a sample containing KTOF$_4$ molecules, the dark stripes in the TEM textures (marked with white arrows in Figure 4(a)) correspond to Kegging clusters (molar mass of Keggin ~ 2900 g/mol), while the narrow dark stripes (marked with yellow arrows in Figure 3(a)) correspond to the OF (molar mass of four OF ~ 660 g/mol). The bright areas represent the flexible chains and the background, Figure 4(a).

The TEM textures and the transmitted electron intensity profile along the layer normal (AB line in Figure 4(a)) exhibit periodicity between 4.5-4.9 nm, Figure 4(b). The fast Fourier transform (FFT) pattern obtained from the texture in Figure 4(a) is shown in Figure 4(c). The FFT pattern shows peaks at 4.8 nm, and 2.4 nm along the layer normal. The periodic peak at 4.8 nm corresponds to the repeated layer distance observed in TEM textures and the transmitted electron intensity profile. The peak at 2.4 nm is the second-order peak, indicating a well-defined modulation.

The transmitted electron intensity profiles along a dark layer (CD line in Figure 4(a)) and a bright layer (EF line in Figure 4(a)) do not show any well-defined periodic intensity peaks, Figure 4(d & e). The absence of a periodic peak along the layer in the FFT pattern confirms the absence of periodicity within the layers. Thus, the KTOF$_4$ self-assembled structures exhibit isotropic in-plane ordering, which is similar to periodic layers of a SmA. If there is a long-range ordering within the layers, as observed in cholesteric and SmC phases [33,35,36], one would expect the formation of disclination lines connecting the center of the concentric spherical structure to the surface of the spherical inclusion. However, the absence of radial disclination lines in the concentric spherical packing of KTOF$_4$, Figure 2(d-g), supports the conclusion that the lamellar layers correspond to the SmA type of ordering. A proposed molecular packing within the periodic layers in Figure 4(f) is schematized in Figure 4(g).

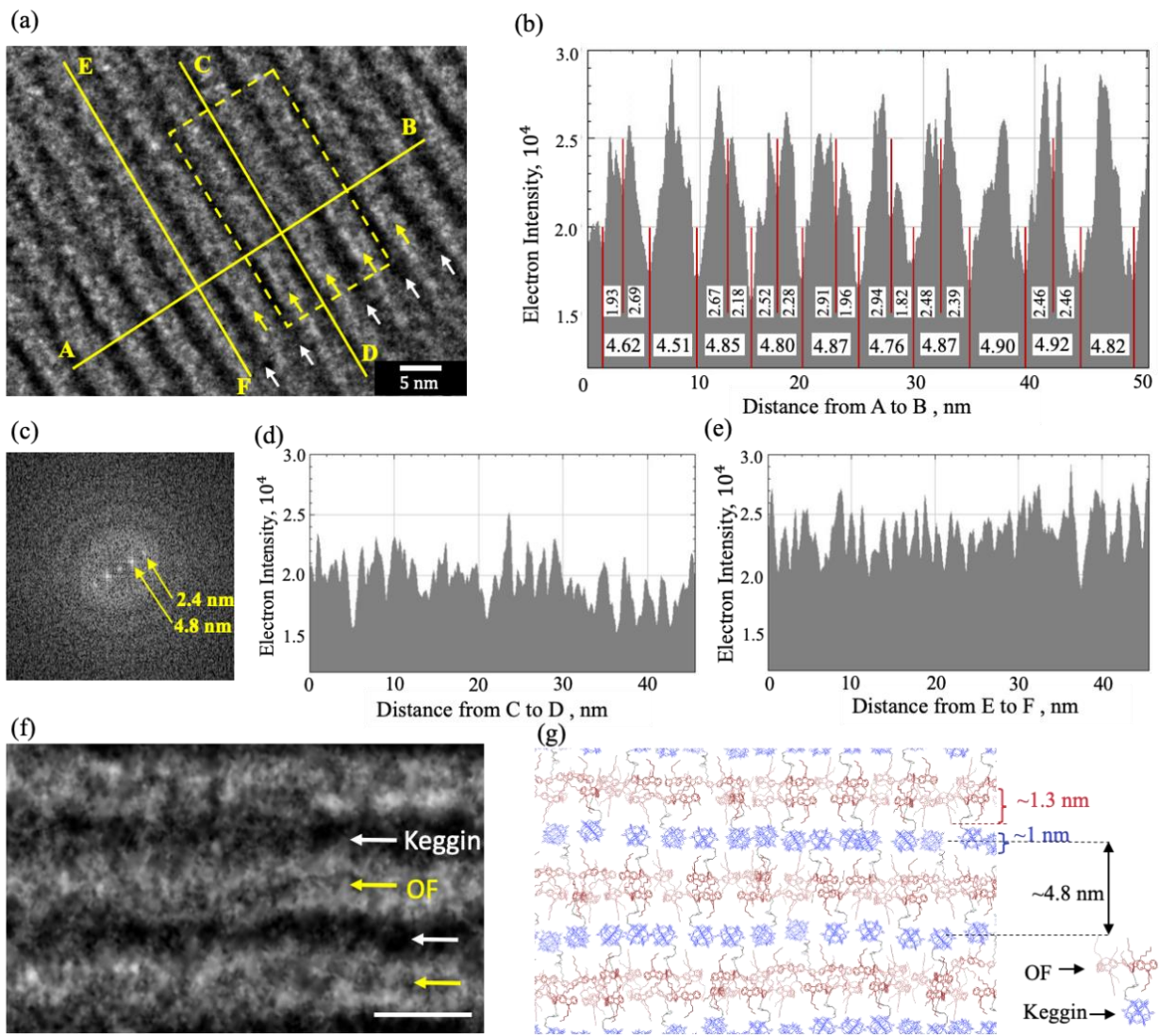

**Figure 4.** (a) Zoomed-in TEM texture of 0.2 mg/ml KTOF$_4$ in 30 vol% dioxane. White arrows marking the dark stripes correspond to Kegging clusters, while yellow arrows marking the narrow dark stripes correspond to the OFs. (b, d & e) The transmitted electron intensity profile along the lines AB, CD, and EF respectively, (c) fast Fourier transform pattern obtained from the texture in 4(a), (f) a zoomed-in section (marked in dashed yellow in 4(a)) of a structure. The scale bar is 5 nm, and (g) proposed molecular packing within the layers.

Periodically modulated LC phases such as SmA exhibit defects of the layered structure such as edge and screw dislocations [51-55]. The dislocations are characterized by Burgers vector, $\boldsymbol{b} = nd\boldsymbol{v}$ where $n$ is an integer, $d$ is the repeat layer distance, and $\boldsymbol{v}$ is the unit vector normal to the layers. According to de Gennes [56] and Kleman and Williams [57], the energy per unit length of an edge dislocation is $W_e = K_1 b^2/2\lambda r_c + w_c$, where $K_1$ is the splay elastic constant, $b = nd$, $\lambda = (K_1/B)^{1/2}$ is the penetration length of material, $B$ is a compressibility modulus, $w_c$ and $r_c$ are the energy and the radius of the core, respectively. Since the energy varies as the square of the Burgers vector, the dislocations of a small Burgers vector ($n = 1$) are common [58]. Allain and Kleman [59] and Kleman [60] showed the core extension of edge dislocation along the Burgers vector is of the order of $\xi_z = d^2/\lambda$. In our case, $\xi_z$ is of the order of $d$, and the dislocation core has a similar extension along the Burgers vector and along the direction perpendicular to it; such a core is often called an isotropic core [60]. The isotropic cores of edge dislocations indicate that for the KTOF4 $\lambda \sim d$. Figure 5(a) illustrates the TEM textures of self-assembled KTOF4 inclusions with edge dislocations indicated by dashed yellow lines. The edge dislocations with small Burgers vectors, $b = d$, are the most frequently observed dislocations in KTOF4 structures. The additional layer of the edge dislocations is always the dark layer in the TEM textures, which corresponds to the hydrophilic part of the KTOF4 molecule (the Keggin cluster); the covalently connected OF rods make a U-turn around this extra layer of Keggin clusters. The core thus can be represented as a pair of disclinations in the OF sublayers, of a strength $+1/2$ around the Keggin clusters and of a strength $-1/2$ in the neighboring region.

In a screw dislocation, the layers are arranged as a spiral staircase [61]. The Burgers vector, $\boldsymbol{b}$, is parallel to the dislocation line. Kleman [58] calculated the energy of screw dislocations is $W_s = Bb^4/128(r_c^{-2} - R^{-2}) + w_c$, where $R$ is the external radius of the sample. The $b^4$ term in energy ensures the stability of screw dislocations with small Burgers vectors. Figure 5(b) illustrates self-assembled KTOF4 inclusions, with screw dislocations indicated by dashed yellow lines. The layers on the two sides of a dislocation line are shifted with respect to each other, by $d/2$, which implies $b = d$. Both edge and screw dislocations are observed in the flat layer structure (formed in 20, 30 and 50-60 vol% dioxane) while dislocations are rarely observed in the concentric spherical layer structure (formed in 35-50 vol% of dioxane).

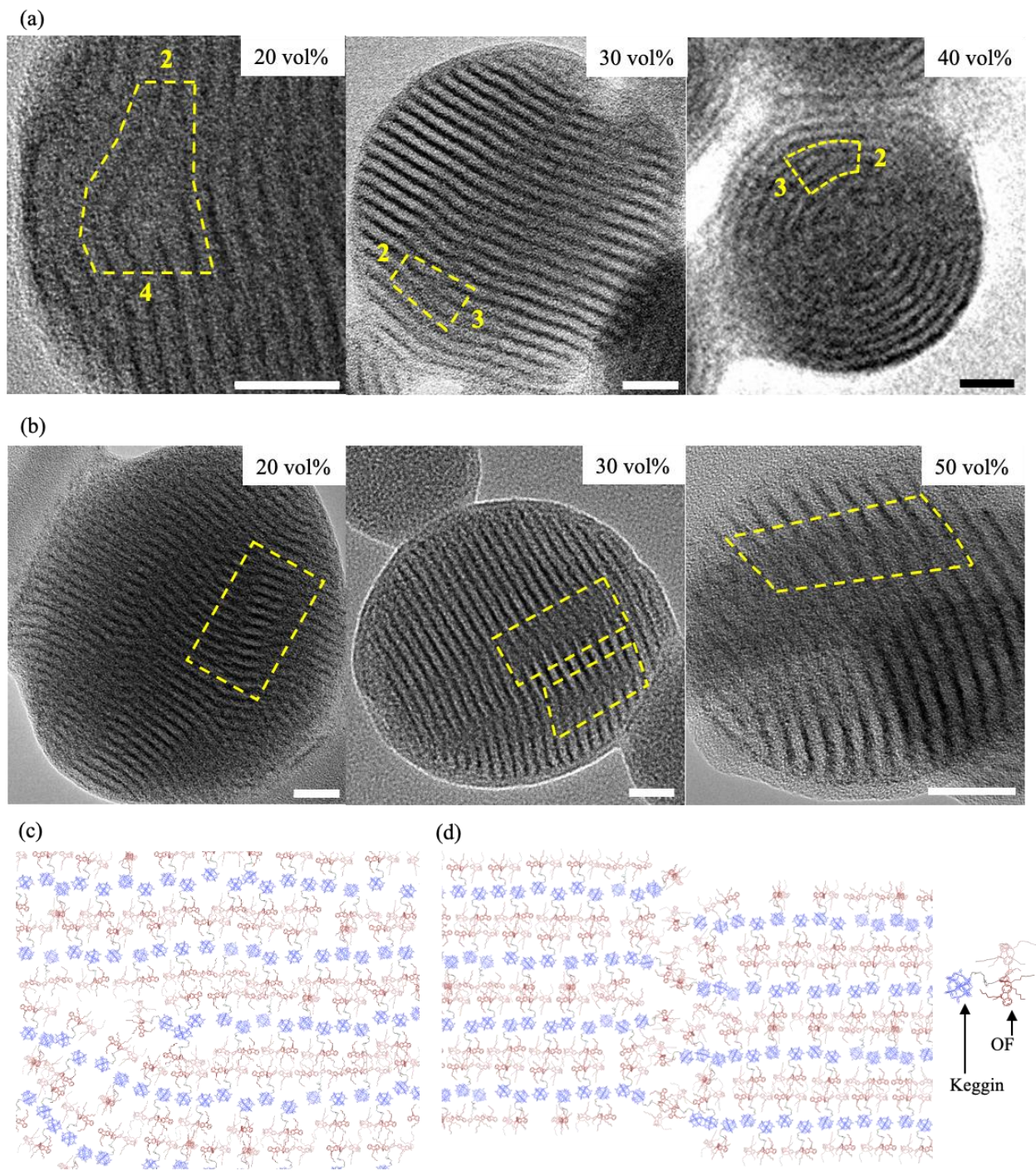

**Figure 5.** Dislocations in self-assembled KTOF$_4$ structures. TEM textures showing (a) edge dislocations, and (b) screw dislocations within the self-assembled layers of 0.2 mg/ml KTOF$_4$. The dioxane content is indicated in the top-right corner of the respective texture. The scale bar is 20 nm for all textures in Figures (a) & (b). X vol% dioxane refers to a solution with a dioxane: water

ratio of X:(100-X). Proposed molecular packing within the layers at the defect site of (c) an edge dislocation, and (d) a screw dislocation.

The explored KTOF$_4$ 2.0 mg/ml dispersion in 50 vol% dioxane shows an intriguing mesomorphic behavior when the solvent is slowly evaporated. Evaporation produces a birefringent fluid residual with birefringence colors and POM textures with extinction brushes resembling those of uniaxial nematic liquid crystals, Figure 6(a). The nematic-like LC$_{KTOF4}$ material aligns perfectly on a rubbed PI2555 layer on a glass substrate, Figure 6(b). An apparent reason for the alignment is orientation of rod-like OF parts along the grooves of the polyimide substrate. The LC$_{KTOF4}$ material shows a biphasic region similar to that of other nematics. Namely, upon cooling, the nematic nuclei appear at 34.5 °C, while at 33.0 °C one observes a homogeneous nematic bulk structure, Figure 6(c). The birefringence of LC$_{KTOF4}$ is measured using a wedge cell method as explained in ref [47,48] and the measured birefringence of the material is high, 0.2.

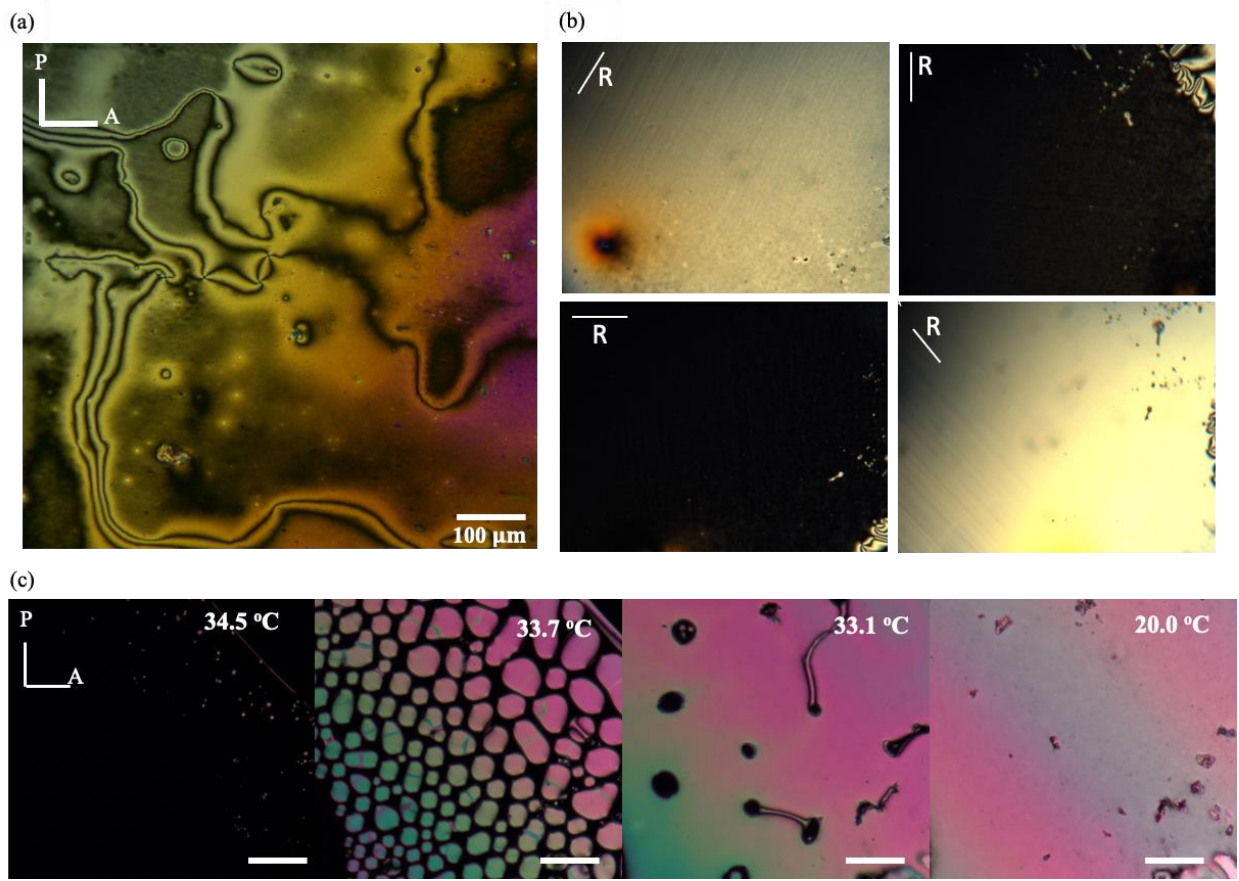

**Figure 6.** Polarizing optical microscopy (POM) study of LC$_{KTOF4}$. LC$_{KTOF4}$ is obtained by allowing 2.0 mg/ml KTOF$_4$ with 50 vol% dioxane to slowly evaporate. POM image of a drop of LC$_{KTOF4}$

(a) on a glass plate, and (b) on the PI2555 coated, rubbed glass plate. The sample is rotated to have different alignments of the rubbing direction, R, with respect to P and A. (c) LC$_{KTOF4}$ progression of phase transition from Isotropic to Nematic while cooling. P and A indicate the polarizer and the analyzer directions and stay the same for all the images. The scale bar is 100 μm.

## 4. Discussion

The study demonstrates that KTOF$_4$ molecules in mixed diaxane-water solutions self-assemble into spherical inclusions with a periodic inner structures. FFT and transmitted electron intensity profiles reveal a well-defined periodicity along the layer normal, Figure 4. The measured period of 4.8 nm in FFT represents the bilayer spacing, Figure 4. The spherical inclusions reveal two types of packings: spherical concentric arrangements of layers and flat layers. The observed textures of spherical concentric packings show no radial disclinations, which signals that there is no long-range orientational order within the layers. In other words, KTOF$_4$ forms SmA type of layered structures in mixed dioxane-water solutions.

The inner structure of LC droplets suspended in an isotropic fluid is controlled by a balance of bulk elasticity, isotropic and anisotropic interfacial interactions [62-65]. To describe the structure, we introduce an apolar unit vector $\hat{\mathbf{n}} \equiv -\hat{\mathbf{n}}$ which is normal to the layers, and a unit vector $\hat{\mathbf{v}}$, which is normal to the spherical surface of the condensed inclusions. The interfacial surface energy can be written as, $F_{int} = 4\pi\sigma_0 R^2$ for a spherical inclusion of radius $R$, where $\sigma_0$ is the interfacial tension [63]. The bulk energy of the sphere can be written as the elastic energy of the spherically bent layers, $F_{ela} = 8\pi K R$, where $K$ is the splay elastic constant for the deformations of $\hat{\mathbf{n}}$ [62,63]. For the tangential alignment of layers at the interface, i.e., when $\hat{\mathbf{n}} \parallel \hat{\mathbf{v}}$, surface anchoring energy at the boundary of the inclusion is, $F_{anch} = \frac{4}{3}\pi W R^2$, where $W$ is the surface anchoring coefficient [62]. The balance between the $F_{int}$ and $F_{ela}$ determines the shape of the anisotropic droplet in isotropic surroundings. Typical values for liquid crystalline cyanobiphenyls in glycerin are $\sigma_0 \sim (10^{-3} - 10^{-2})$ Jm$^{-2}$, $W \sim (10^{-6} - 10^{-5})$ Jm$^{-2}$ and $K = 10^{-11}$ N [66]. At the nematic-isotropic interface of 5CB, $\sigma_0 \approx 2 \times 10^{-5}$ Jm$^{-2}$ and $W \approx 5 \times 10^{-7}$ Jm$^{-2}$ [67]. As a result, any macroscopic LC droplet larger than $K/\sigma_0 \approx (1-10)$ nm should be spherical. In our study, we observe that the self-assembled inclusions of KTOF$_4$ in 20–60 vol% dioxane are spherical with radii 25-90 nm, Figure 2. This observation suggests that in the explored system, $2K/\sigma_0 < 25$ nm. If one assumes $K = 10^{-11}$ N, then $\sigma_0 > 8 \times 10^{-4}$ Jm$^{-2}$.

Anisotropic surface interaction plays a critical role in determining the layer structure inside the LC droplets. For an LC droplet with $R \ll 6K/W$, spatial variations of the director $\hat{\mathbf{n}}$ are avoided, and $\hat{\mathbf{n}}(r)$=constant is maintained at the expense of violating boundary conditions. Conversely, droplets with $R \gg 6K/W$ satisfy boundary conditions by aligning molecules along the easy axis [62,63]. For KTOF$_4$ inclusions, both flat and spherical layer packings are observed in spheres of 25-90 nm. The type of layer does not depend on the size of the inclusion, which signals that the type of packing is affected by the subtle balance of bulk elasticity and surface anchoring. Stiff layers would prefer to form flat packings, while soft flexible layers would form concentric spheres. This feature can explain the observation for solutions with low and medium dioxane content, as discussed below.

The spherical packings in the medium dioxane content can be explained simply by the different affinity of the polar Keggin clusters and nonpolar OF rods to the solvent. If the solvent prefers to interface with one but not the other, the resulting packing will be spherical. If the concentration of dioxane decreases and the polarity of the solvent increases, then the hydrophobic attractions between OF rods enhance. To minimize the area of contact with a polar solvent, the rods would tend to be parallel to each other, Figure 6. The layers would become stiffer and thus less prone to follow the curvature of the confining volume. This is in line with the observation of flat layers at low dioxane concentrations.

The observation of flat layers at the high content of dioxane is less clear. Tentatively, the non-polar solvent allows the OF rods to form strongly curved U-turn regions around the layers of Keggin clusters, Figure 7(a), as in the cores of edge dislocations, Figure 5(c). These U-turns are not very costly energetically since the hydrophobic rods are not aligned to be strictly parallel to each other to minimize the contact with the solvent. These U-tern regions might form at the interface of the layers with the non-polar solvent, thus shielding the polar Keggin clusters from contact with it. Of course, this scenario is likely to be complicated by the electrostatic forces that likely change with the solvent polarity; a more detailed analysis requires further experiments.

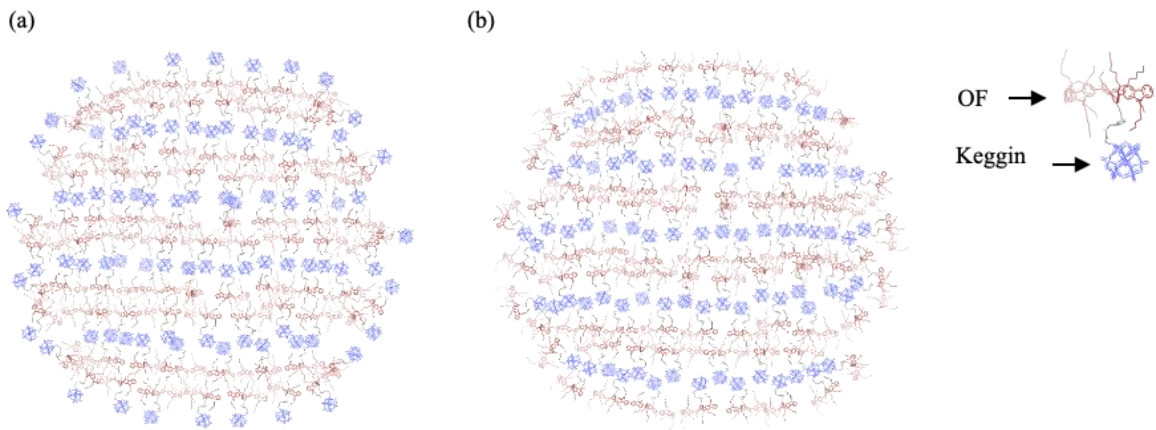

**Figure 7.** Proposed KTOF$_4$ molecular packing within the flat layers with (a) Keggin, and (b) OF as surface layer composition.

Finally, we observe an unexpected phenomenon of formation of a nematic-like structure in KTOF$_4$ dispersions when the solvent is slowly evaporated. The residual material is fluid, highly birefringent and exhibits typical nematic textures, Figure 6(a). Unfortunately, attempts to prepare a similar nematic phase by producing highly concentrated KTOF$_4$ dispersions did not produce a nematic phase; apparently, the process of self-organization is complex and requires unusual steps such as slow evaporation of the solvent. The solvent-free KTOF$_4$ material, a solid in powder form, undergoes a transition to an unspecified weekly birefringent phase at 200 °C upon heating. Upon cooling, the material crystallizes at 160 °C. Further studies are needed to explore the strongly concentrated solutions and solvent-free samples of KTOF$_4$.

## 5. Conclusions

In this work, we study the self-assembled structures of KTOF$_4$ amphiphilic molecules in dioxane and water-mixed solutions. 0.2 mg/ml KTOF$_4$ molecules form self-assembled layered structures of a smectic A type with sublayers of the Keggin and OF of the molecules that segregate away from each other. The self-assembled spherical inclusions show either concentric packing of the layers, when the concentration of dioxane is in the range 35-50 vol% or flat layers when the dioxane concentration is higher or lower. The electron microscopy textures demonstrate elementary dislocations of the edge and screw type. The organization of molecules within the layers shows no long-range orientational order as we do not observe radial disclinations in the spherical packings. In other words, the self-assembled lyotropic phases are of the SmA type. Furthermore, increasing the KTOF$_4$ concentration through slow evaporation leads to the formation of a birefringent bulk phase of the nematic type. This study provides an insight into a rich morphogenesis of self-assembly of KTOF$_4$ amphiphilic molecules of nontrivial shape in mixed dioxane-water solutions.

**Author Contributions:** N.P.H. conducted the experiments, analyzed and discussed the data, and contributed to the writing, Y.Z & J.W synthesized the material (KTOF$_4$) and prepared samples. M.G. guided the TEM imaging and image analysis. T.L. directed research for the portion at The University of Akron. O.D.L. contributed to the writing and directed the research. All authors have read and agreed to the published version of the manuscript.

**Funding:** O.D.L. acknowledges support from the NSF DMR-2215191. T.L. acknowledges support from the NSF DMR2215190 and The University of Akron.

**Data Availability Statement:** The datasets generated during and/or analyzed during the current study are available from the corresponding author upon reasonable request.

**Conflicts of Interest:** The authors declare no conflicts of interest.

**Acknowledgments:** The TEM imaging was carried out at AMLCI Characterization Facility at Kent State University.